\documentclass[
					aps, 							
					pre,							
					twocolumn,    				
					10pt,
					showpacs, 				
					preprintnumbers,		
					amsmath,
					amssymb,					
					english]{revtex4-1}

\usepackage[english]{babel}
\usepackage{bm}		
\usepackage[unicode, pdfstartview= FitH]{hyperref}

\begin{document}

\preprint{OS.R82/08-en}

\title{Open statistical ensemble: new properties (scale invariance, application to small systems, meaning of surface particles, etc.)}

\author{V. M. Zaskulnikov}
\homepage{http://www.zaskulnikov.ru}
\email[]{zaskulnikov@gmail.com }

\affiliation{
Institute of Chemical Kinetics and Combustion, Institutskaya, 3, Novosibirsk, 630090, Russian Federation
}

\date{\today}

\begin{abstract}

A new statistical ensemble is examined using the example of classical one-component simple fluid. It's logical to call it an open ensemble, because its peculiarity is the inclusion in the consideration some surrounding area. Calculations point to the necessity of taking into account the restricting surface, exactly when the system is not separated by anything from the bath, and the whole medium is uniform. 

The ``surface tension coefficient'', included in the partition function corresponds to the interface of the fluid and hard solid, due to the strict compliance of probability and potential limitations. The number of surface particles corresponds exactly to near surface number density distortions (oscillations) arising in the neighborhood of fluctuation cavities.

In contrast to grand canonical ensemble, an open statistical ensemble satisfies the scale invariance requirement: general term of the included subsystem distribution corresponds to that of the original system. 

It is this ensemble which should be used where consideration of a truly open system is required, since it properly integrates the surface terms. Furthermore, this ensemble may be employed in studies of small systems, since it has no lower limits for the volume of the system. Finally, it is useful in the investigation of fluctuations. For example, it demonstrates that the variance (the mean square deviation) of the number of particles is divided into the bulk and surface terms.

\end{abstract}


\maketitle

\section{\label{sec:01}Introduction}

The Gibbs grand canonical distribution is one of the foundations of statistical physics and, in its basic parts, is not doubt. However, it will be interesting to check the grand canonical ensemble (GCE) relations for correspondence to some criteria. One of such criteria is a peculiar kind of scaling, which lies in the fact that the distribution for the subsystem (calculated from the source) should functionally coincide with the original.

Such testing seemed reasonable from the very beginning. Indeed, the familiar treatment of nucleation dating back to J.W.Gibbs  \cite[p.242]{gibbs1961} involves nucleus surface into consideration. The grand canonical distribution itself describes the fluctuations of the particles number in a given volume. Thus, initially, at the GCE level, for example, the probability of the formation of the hole must include the surface terms. 

The presence of such a term in the partition function of GCE was established earlier \cite{Bellemans1962}, however, it is easy to see (sections \ref{subsec:03a_1}, \ref{subsec:04f}, \ref{subsec:06b}), that it is included with the wrong sign.

Consequently, the conventional procedure of passing to the thermodynamic limit is in doubt. The thing is that the increase in the GCE volume does not eliminate the boundary distortions because they remain ``glued'' to its borders. Correct thermodynamic limit must fix the volume of observation, and take away the boundary distortions to infinity  (section \ref{subsec:06a}).

Introduction of terms defining the interaction over the surrounding surface in statistical distribution calls for the exploration of the medium beyond the limits of the system. Thus we arrive at the idea of an open statistical ensemble (OSE). It can be constructed with the use of the scaling check procedure.

Though OSE construction itself is connected with the violation of GCE scale invariance, it is preserved during the subsequent breakdown of the system.

In the case of one-component system liquid/gas the surface tension coefficient depends solely on one variable, either pressure or temperature, due to the phase equilibrium curve. The situation under discussion is beyond the limits of this condition, and ``surface tension coefficient'' depends on two variables. As a result, a non-zero number of surface particles arises for the system boundary. It is necessary to establish the physical meaning of this parameter.

Another issue complicating and limiting the work with the GCE is the impossibility of direct application of this distribution to small systems. This limitation is  habitual and, at the first glance, seems reasonable but it has no deep foundation. There is no reason not to build the distribution, which will be valid for arbitrarily large and arbitrarily small volumes. In particular, this would allow one to calculate the number density fluctuations for small volumes.

In addition, it is interesting to simplify an extremely complicated method of calculations of surface terms, based on the diagram technique \cite{Bellemans1962}. It is impossible to work with a series if you do not have the general term.

It should be emphasized that in the given case the surface terms appear in the description of a homogeneous medium involving no real surface but just a hypothetical one, contouring the system under study.

\section{\label{sec:02}Primary definitions}

To simplify  operations, in most cases we shall define the integration domain by characteristic functions 
\begin{equation}
\psi^v(\bm{r}_i) = \psi^v_i = 
	 \left\{ 
			\begin{array}{ll} 
         1 & (\bm{r}_i \in v)\\   
         0 & (\bm{r}_i \notin v)
     	\end{array}  
		\right.
		\label{equ:01}
\end{equation}

and

\begin{equation}
\chi^v_i = 1 - \psi^v_i = 
	 \left\{ 
			\begin{array}{ll} 
         0 & (\bm{r}_i \in v)\\   
         1 & (\bm{r}_i \notin v),
     	\end{array}  
		\right.
		\label{equ:02}
\end{equation}
where $v$ denotes the volume of the domain, and integration in all the integrals is considered to be performed over the infinite space unless otherwise specified. 

Note that the algebra of characteristic functions plays an important part in the approach at hand. Besides, as it will be seen further, the role of these functions is significant from the standpoint of distribution functions determination.

\subsection{\label{subsec:02a}Canonical ensemble}

The probability density to find the given spatial configuration of a specific set of particles  \cite[p.181]{hillstatmeh1987} is defined by the expression
\begin{equation}
P^{(k)}_{1...k} = \frac{1}{Z_N^V}\int \left [ \prod_{l = k+1}^N \psi^V_l \right ] \exp(-\beta U^N_{1...N})d\bm{r}_{k+1}...d\bm{r}_N,
\label{equ:03}
\end{equation}
where $N$ is the number of particles in the system, $\beta = 1/k_BT$, $k_B$ is the Boltzmann constant, $T$ is the temperature, $U^N_{1...N}$ is the interaction energy of particles between each other, $V$ is the system volume. Integration is performed over the coordinates of the ensemble particles  $\bm{r}_{k+1}...\bm{r}_N$.

$Z_N^V$ is the configuration integral
\begin{equation}
Z_N^V = \int \left [ \prod_{k = 1}^N \psi^V_k \right ]  \exp(-\beta U^N_{1...N}) d\bm{r}_1...d\bm{r}_N. 
\label{equ:04}
\end{equation}

Passing to the distribution functions for arbitrary set of particles, we have
\begin{equation}
\varrho^{(k)}_{C,1...k} = \frac{N!}{(N-k)!}P^{(k)}_{1...k}, 
\label{equ:05}
\end{equation}
where $\varrho^{(k)}_{C,1...k}$ gives the probability density to find a given configuration of $k$ arbitrary particles for the canonical ensemble.

\subsection{\label{subsec:02b}Grand canonical ensemble}

Average equality (\ref{equ:05}) over fluctuations of the number of particles, i.e., apply the operation $\sum_{N=0}^\infty P_N^V$ to both its sides, where 
\begin{equation}
P_N^V = \frac{z^N Z_N^V}{N!\Xi_V} 
\label{equ:06}
\end{equation}
is the probability for GCE to have a definite number of particles $N$ inside the volume $V$. Here $z$ is the activity
\begin{equation}
z = \frac{e^{\mu/k_BT}}{\Lambda^3}, 
\label{equ:07}
\end{equation}
where $\mu$ is the chemical potential, $\Lambda = h/\sqrt[]{2 \pi mk_BT}$, $h$ is the Planck constant, $m$ is the particle mass, and $\Xi_V$ is a large partition function of the system of the volume $V$
\begin{equation}
\Xi_V = 1 + \sum_{N=1}^\infty \frac{z^N Z_N^V}{N!}. 
\label{equ:08}
\end{equation}

Thus
\begin{equation}
\varrho^{(k)}_{G,1...k} = \sum_{N=k}^\infty \varrho^{(k)}_{C,1...k} P_N^V,
\label{equ:09}
\end{equation}
or 
\begin{eqnarray}
&&\varrho^{(k)}_{G,1...k} (\psi^V) = \frac{z^k}{\Xi_V} \left \{ \exp(-\beta U^{k}_{1...k})  +  \sum_{N=1}^\infty \frac{z^N}{N!} \right . \label{equ:10} \\
&& ~~~~~~ \left . \times  \int   \left [ \prod_{l = k+1}^{k+N} \psi^V_l \right ]  \exp(-\beta U^{N+k}_{1...N+k})d\bm{r}_{k+1}...d\bm{r}_{k+N} \right \},  \nonumber
\end{eqnarray}
 
where $\varrho^{(k)}_{G,1...k}(\psi^V)$ is the $k$-particle distribution function for GCE. As already mentioned, these functions specify the probability density to find a certain configuration of arbitrary particles. For ideal gas $\varrho^{(k)}_{G,1...k} = \varrho^{k}$, where $\varrho = \overline{N}/V$ is the number density. 

Here $\psi^V$ cannot be treated as the definition domain of the function in the ordinary sense, since, as it will be seen further, the coordinates of free particles can fall outside the limits it defines. For brevity, we shall call it the assignment domain. For example, in equation (\ref{equ:10}) $\psi^V$ is the assignment domain of the function $\varrho^{(k)}_{G,1...k}$.

Characteristic functions appear both in the nominator and in the denominator of (\ref{equ:10}); in the last case through $\Xi_V$ (\ref{equ:08}). This will be discussed in detail below. 

Further the significance of the assignment domain $\psi^V$ will become obvious, so we shall give it in the explicit form, and the fact that $\varrho^{(k)}_{G,1...k}(\psi^V)$ belongs to GCE type will be denoted by the index $G$.

Commonly, the assignment domain $\psi^V$ is noted and controlled only slightly, however, from the standpoint of the present paper it is of crucial importance.

Complete integrals $\varrho^{(k)}_{G,1...k}(\psi^V)$ satisfy the relation
\begin{equation}
\int \left [ \prod_{i = 1}^k \psi^V_i \right ] {\varrho}^{(k)}_{G,1...k} (\psi^V) d\bm{r}_1... d\bm{r}_k = \left \langle  \frac{N!}{(N-k)!} \right \rangle,
\label{equ:11}
\end{equation}
following from the definition. Here the triangular brackets denote averaging over the number of particles of type (\ref{equ:09}). Note that for GCE integration and assignment domains in (\ref{equ:11}) rigorously coincide.

\subsection{\label{subsec:02c}Presence of an external field}

The configuration integral of the inhomogeneous closed system is given by the expression
\begin{equation}
Z^U_N = \int\limits_{V}^{} \exp(-\beta\sum_{i=1}^N u_i-\beta U^N_{1...N}) d\bm{r}_1...d\bm{r}_N, \\
\label{equ:12}
\end{equation}
where $u_i$ is the energy of the interaction of the $i$-th particle with the force field.

For the GCE, we introduce the quantity
\begin{equation}
\Xi^U_V = 1+ \sum_{N=1}^\infty \frac{z^N Z^U_N}{N!}, 
\label{equ:13}
\end{equation}
which is obviously the large partition function of the system in the presence of an external field.

\section{\label{sec:03}OSE construction}
 
The idea of OSE construction consists in the separation of a certain subsystem rather far removed from the boundaries of the whole GCE system. In practice it is sufficient if the subsystem is smaller than the system just by several atomic layers. Lower bounds of the system size are absent, so however small open ensembles may be considered.

\subsection{\label{subsec:03a}OSE partition function}

The probability of a hole formation is an important parameter that defines the ensemble partition function. Thus we start with the examination of the first and the most significant term of this distribution, and then pass to other terms. First, the canonical ensemble will be considered, then GCE, and finally OSE definition.

Consider the probability of the fluctuation formation of the cavity inside a uniform statistical system - the region of the volume $v$ containing no particles. Let us take that the cavity is rather far away from the system boundaries.

Evidently, the desired probability for the closed system is defined as
\begin{equation}
p^v_{C,0} = \frac{1}{Z_N^V}\int \left [ \prod_{k = 1}^N(\psi^V_k-\psi^v_k) \right ] \exp(-\beta U^N_{1...N}) d\bm{r}_1...d\bm{r}_N
\label{equ:14}  
\end{equation}
or
\begin{equation}
p^v_{C,0} = \frac{Z_N^{V-v}}{Z_N^V}.
\label{equ:15}  
\end{equation}

Here the index $C$ on $p^v_{C,0}$ shows that the probability is defined in the canonical ensemble on condition that $N$ particles are in the volume $V$. 

Passing to the probability for GCE
\begin{equation}
p^v_{G,0} = \sum_{N=0}^\infty p^v_{C,0} P_N^V,
\label{equ:16}
\end{equation}
and substituting (\ref{equ:06}) in (\ref{equ:16}) we obtain
\begin{equation}
p^v_{G,0} = \frac{\Xi_{V-v}}{\Xi_V}.
\label{equ:17}
\end{equation}

As is seen from the derivation procedure, both large partition functions are considered at one and the same activity.

Introduce the quantity
\begin{equation}
\Upsilon_v =  \frac{\Xi_V}{\Xi_{V-v}},
\label{equ:18}  
\end{equation}
by definition it is the partition function of OSE system of the volume $v$. 

Strictly speaking, it would be reasonable to retain the index $V$ in the partition function $\Upsilon_v$ as well, or to define it via passing to the limit $V\rightarrow \infty$. This passage will be done in examining the expansions in powers of the activity. However, we shall see that the dependence on the system volume $V$ is so slight that (\ref{equ:18}) is a rather good definition even when the boundaries of two systems are separated just by several atomic layers.

Essentially, the difference between OSE and GCE is in surface effects; this is clear from (\ref{equ:18}). When the subsystem does not interact with the medium, $\Xi_V = \Xi_v \Xi_{V-v}$, and this equation reduces to a banal one
\begin{equation}
\Upsilon_v = \frac{\Xi_v \Xi_{V-v}}{\Xi_{V-v}}  = \Xi_v,
\label{equ:19}  
\end{equation}
thus demonstrating the identity of the ensembles.

\subsection{\label{subsec:03a_1}Equivalence of the potential and statistical limitations}

Potential and statistical restrictions closely correspond to each other for the potential of hard solid. This is evident from the fact that the Bolzmann factor corresponds to the characteristic function of (\ref{equ:01}) type in this case. Consequently, the partition function (\ref{equ:12}) coincides exactly with $Z_N^{V-v}$, appearing in (\ref{equ:15}), and (\ref{equ:13}) - with $\Xi_{V-v}$ from (\ref{equ:17}).

Thus, equations (\ref{equ:15}), (\ref{equ:17}) may be interpreted as the ratios of partition functions where the denominator involves the partition function of the initial system, and the nominator - that of the system with hard solid immersed in it. Therefore the logarithm of the ratio of partition functions acquires (accurate to the factor) the meaning of the change in $\Omega$ - potential (grand potential)

\begin{eqnarray}
\Omega^u - \Omega &=& [-P(V-v) + \sigma a] - [-PV] \nonumber \\
 &=& vP(z,T)  +a \sigma (z,T). \label{equ:20}  
\end{eqnarray}

Here the index $u$ denotes the system with the immersed solid, $P(z,T)$ is the pressure, $a$ - the surface restricting the subsystem (hard solid), and $\sigma (z,T)$ - ``surface tension coefficient'' at the hard solid/fluid interface. 

Using (\ref{equ:20}), obtain
\begin{equation}
\Upsilon_v = \exp{\beta [ vP(z,T)  + a\sigma (z,T) ]}.
\label{equ:21}
\end{equation}
This expression is valid, obviously, for rather smooth surface only, i.e., for not very small volume $v$.

Naturally, 
\begin{equation}
p^v_0 =   \frac{1}{\Upsilon_v},
\label{equ:22}  
\end{equation}
where $p^v_0$ is the probability of a hole of the volume $v$ formation for OSE.

So, the ``surface tension coefficient'' included in the OSE always corresponds to the boundary between fluid and hard solid. 

The near surface number density oscillations close to the above fluctuation cavities also correspond exactly to the distortions at the boundary of hard solid and fluid. This also agrees with the Bolzmann idea concerning fluctuation thermodynamics (entropy) definition in terms of appropriate quantities for the system in a such field, where a given fluctuation is an equilibrium configuration.

Below we shall show the separation of surface terms in OSE, here we also refer to the results of the work \cite{Bellemans1962} from which it is clear that for GCE 
\begin{equation}
\Xi_V = \exp{\beta [ VP(z,T)  - A\sigma (z,T) ]},
\label{equ:23}
\end{equation}
where $A$ is the surface restricting the system. 

Substituting (\ref{equ:23}) in (\ref{equ:18}) it is easy to see that the expressions (\ref{equ:21}) and (\ref{equ:23}) correspond to each other. Quite interesting are the opposite signs of surface terms in GCE and OSE. In section \ref{subsec:06b} a comparison of these ensembles will be made.

\subsection{\label{subsec:03b}OSE distribution}

Let us calculate the general term of OSE distribution. Thus, consider the probability for a uniform system to find exactly $m$ particles in the volume $v$ - $p^v_m$. In essence, calculations will correspond to the above calculation of a zero term. We begin with the probability for the canonical ensemble

\begin{eqnarray}
p^v_{C,m} = \binom{N}{m} && \frac{1}{Z_N^V}\int\exp(-\beta U^N_{1...N})\label{equ:24}  \\
 \times &&  \left [\prod_{i = 1}^m\prod_{j = m+1}^N\psi^v_i (\psi^V_j-\psi^v_j) \right ]  d\bm{r}_1...d\bm{r}_N. \nonumber
\end{eqnarray}

Then we pass to analogous probability for GCE
\begin{equation}
p^v_{G,m} = \sum_{N=m}^\infty p^v_{C,m} P_N^V
\label{equ:25},
\end{equation}
and substituting (\ref{equ:06}) gives 

\begin{eqnarray}
p^v_{G,m}& = & \frac{z^m}{m!\Xi_V}\sum_{N=0}^\infty \frac{z^N}{N!}\int\exp(-\beta U^{N+m}_{1...N+m}) \nonumber \\
 &\times&  \left [\prod_{i = 1}^m\prod_{j = m+1}^{m+N}\psi^v_i (\psi^V_j-\psi^v_j) \right ]  d\bm{r}_1...d\bm{r}_{N+m}. \label{equ:26}
\end{eqnarray}

Eventually, by multiplication and division of (\ref{equ:26}) by $\Xi_{V-v}$ and taking the sum of the series, we get
\begin{equation}
p^v_m = \frac{1}{m! \Upsilon_v} \int \left [ \prod_{i = 1}^m \psi^v_i \right ] \varrho^{(m)}_{G,1...m}(\psi^V-\psi^v) d\bm{r}_1...d\bm{r}_m. 
\label{equ:27}
\end{equation}

This is just the OSE distribution general term we are interested in. The sense of this expression is clear: the probability of finding $m$ particles in the volume $v$ is determined by the probability of finding the hole of the same size, $1/\Upsilon_v$, multiplied by the probability density of revealing some configuration of $m$ particles (on condition that other particles are beyond the limits of the volume) integrated over all configurations. The factor $1/m!$ arises due to the symmetry (indistinguishability) under particles permutations.

(Here we deal exactly with the case where the coordinates of particles $1,...m$ are beyond the limits of the distribution function assignment domain defined by the functions $\psi^V-\psi^v$.)

First note the similarity between expression (\ref{equ:27}) and the general term of GCE distribution (\ref{equ:06}). They coincide accurate to the normalizing factor in the approximation 
\begin{equation}
\varrho^{(m)}_{G,1...m} \approx z^m \exp(-\beta U^{m}_{1...m}), 
\label{equ:28}
\end{equation}
i.e., in the low density limit when the first term plays the main role in the distribution function expansion \cite[p.410]{hillstatmeh1987}, (\ref{equ:a05}).  

In particular, this means that for low densities the role of surface tension is less important. As will be seen later the reason is that in this case the expansion of surface terms into a series in activity begins with a quadratic term. 

Expressions (\ref{equ:18}), (\ref{equ:21}) and (\ref{equ:27}) solve the problem of OSE construction, and now we can turn to its properties.

\section{\label{sec:04}Some properties of OSE}

\subsection{\label{subsec:04f}$p^v_0$ as the fluctuation probability}

The equation for the cavity formation probability which has the form 
\begin{equation}
p^v_0 = \exp{- \beta [vP(z,T) + a \sigma (z,T)]}
\label{equ:29}
\end{equation}
(as follows from (\ref{equ:21}), (\ref{equ:22})) agrees with the general expression for fluctuation probability  
\begin{equation}
p_f \propto \exp{( -\beta R_{min})},
\label{equ:30}  
\end{equation}
\cite[p.339]{landaulifshitz1985}, where $R_{min}$ is the minimum work of fluctuation removal. 

Indeed, since the chemical potential of particles outside the field is not changed under local imposition of the field, we can take that in this case the process of creation/removal of fluctuation occurs at constant chemical potential and temperature, i.e., natural variables of $\Omega$-potential (grand potential). So minimum cavity formation work is equal to the change in $\Omega$-potential (\ref{equ:20})

\begin{equation}
 R_{min} = \Omega^U - \Omega  = vP(z,T)  +a \sigma (z,T). \label{equ:31}  
\end{equation}

As we are interested in the fluctuation at which the density changes in a stepwise way, this fluctuation creation process must be performed by hard solid (see Section \ref{subsec:03a_1}).

\subsection{\label{subsec:04a}Normalization condition}

Normalization condition
\begin{equation}
\sum_{m=0}^{\infty}  p^v_{G,m} = 1
\label{equ:32}
\end{equation}
gives the expression

\begin{eqnarray}
&&\sum_{m=0}^{\infty} p^v_{G,m} = \frac{1}{\Xi_V} \sum_{m=0}^\infty \frac{1}{m!} \sum_{N=m}^\infty \frac{z^N}{(N-m)!}\label{equ:33}  \\
&&\times \int \left [\prod_{i = 1}^m\prod_{j = m+1}^N\psi^v_i (\psi^V_j-\psi^v_j) \right ] \exp(-\beta U^N_{1...N}) d\bm{r}_1...d\bm{r}_N. \nonumber
\end{eqnarray} 
 
Changing the order of summation, we have

\begin{eqnarray}
&&\sum_{m=0}^{\infty} p^v_{G,m} = \frac{1}{\Xi_V} \sum_{N=0}^\infty \frac{z^N}{N!}\int \sum_{m=0}^N \binom{N}{m}\label{equ:34}  \\
&&~~~\times   \left [\prod_{i = 1}^m\prod_{j = m+1}^N\psi^v_i (\psi^V_j-\psi^v_j) \right ] \exp(-\beta U^N_{1...N}) d\bm{r}_1...d\bm{r}_N. \nonumber
\end{eqnarray} 

Using the Bolzmann factor symmetry under permutations of particles and binomial formula, we arrive at

\begin{eqnarray}
\sum_{m=0}^{\infty} p^v_{G,m} &=& \frac{1}{\Xi_V} \sum_{N=0}^\infty \frac{z^N}{N!} \label{equ:35}  \\
&\times & \int\left [\prod_{i = 1}^N\psi^V_i \right ] \exp(-\beta U^N_{1...N}) d\bm{r}_1...d\bm{r}_N = 1. \nonumber
\end{eqnarray}

\subsection{\label{subsec:04b}The mean number of particles}

With the aid of (\ref{equ:27}), one can find the mean number of particles in a certain volume $v$

\begin{eqnarray}
&&\overline  m = \frac{1}{\Xi_V} \sum_{m=1}^\infty \frac{1}{(m-1)!} \sum_{N=m}^\infty \frac{z^N}{(N-m)!}\label{equ:36}  \\
&&\times \int \left [\prod_{i = 1}^m\prod_{j = m+1}^N\psi^v_i (\psi^V_j-\psi^v_j) \right ] \exp(-\beta U^N_{1...N}) d\bm{r}_1...d\bm{r}_N. \nonumber
\end{eqnarray}

Changing the order of summation and applying binomial distribution and symmetry under permutations again, we have
\begin{equation}
\overline  m =  \int  \psi^v_1  \varrho^{(1)}_{G,1}(\psi_V) d\bm{r}_1. 
\label{equ:37}
\end{equation}

Though relation (\ref{equ:37}) seems banal, it has a non-trivial sense. Since in the given case integration is performed far from the system boundaries, the function $\varrho^{(1)}_{G,1}$ involves no boundary distortions, thus, unlike GCE, the mean number of particles for OSE is exactly equal to the number of volume particles.

So,
\begin{equation}
\overline  m =  m_b, 
\label{equ:38}
\end{equation}
where 
\begin{equation}
m_b = v\frac{\partial P}{\partial \mu},
\label{equ:39}
\end{equation}
\footnote[1]{All differentiations are hereinafter performed with constant volume and temperature values, unless it is stated otherwise.} since in the limits of the volume $v$ the following equation holds
\begin{equation}
\varrho^{(1)}_{G,1}(\psi_V) = \varrho = \frac{\partial P}{\partial \mu}.
\label{equ:40}
\end{equation} 

This will be the subject of further discussion in connection with the comparison between OSE and GCE.

\subsection{\label{subsec:04c}Fluctuations of the number of particles} 

Averaging the expression $m(m-1)(m-2)...(m-k+1)$ with the use of (\ref{equ:27}), we obtain

\begin{eqnarray}
&&\left \langle  \frac{m!}{(m-k)!} \right \rangle = \frac{1}{\Xi_V} \sum_{m=k}^\infty \frac{1}{(m-k)!} \sum_{N=m}^\infty \frac{z^N}{(N-m)!}\label{equ:41}  \\
&&\times \int \left [\prod_{i = 1}^m\prod_{j = m+1}^N\psi^v_i (\psi^V_j-\psi^v_j) \right ] \exp(-\beta U^N_{1...N}) d\bm{r}_1...d\bm{r}_N. \nonumber
\end{eqnarray}

Proceeding as in the case with (\ref{equ:33}) $\rightarrow$ (\ref{equ:34}) $\rightarrow$ (\ref{equ:35}), we have
\begin{equation}
\int \left [ \prod_{i = 1}^k \psi^v_i \right ] \varrho^{(k)}_{G,1...k}(\psi^V) d\bm{r}_1...d\bm{r}_k =  \left \langle  \frac{m!}{(m-k)!} \right \rangle. 
\label{equ:42}
\end{equation}

Though this expression seems similar to (\ref{equ:11}), there is an essential distinction: integration and assignment domains of the functions $\varrho^{(k)}_{G,1...k}$ do not coincide in this case. This results in the following. First, in contrast to the case with GCE, the surface terms enter in (\ref{equ:42}) in the proper way, since here we are not concerned with the region of near surface distortions of $\varrho^{(k)}_{G,1...k}$ at the outer boundary of GCE.

Second, unlike (\ref{equ:11}), in (\ref{equ:42}) the integration domain may be however small, and this allows one to employ this expression to study fluctuations in small volumes.

For example, quadratic fluctuation in the volume $v$ may be found from the equation
\begin{equation}
\int  \psi^v_1 \psi^v_2 \left (  \varrho^{(2)}_{G,1,2} - \varrho^{(1)}_{G,1} \varrho^{(1)}_{G,2} \right ) d\bm{r}_1 d\bm{r}_2 =  \overline {m^2} - {\overline m}^2 - \overline {m}. 
\label{equ:43}
\end{equation}

Further the fluctuations will be discussed in greater detail.

Note that theorems (\ref{equ:32}), (\ref{equ:37}), (\ref{equ:42}) can be proved in a different way, namely,  with the help of a series in powers of the activity \cite{zaskulnikov200911}.

\subsection{\label{subsec:04d}Small systems}

It is clear from the derivation of OSE expressions that this distribution may be applied to however small volumes including those less than a molecule size. However, as already mentioned, the expression for partition function (\ref{equ:21}) does not work for such sizes. For very small $v$ the partition function may be calculated as follows. 

Just as from physical considerations, it follows from (\ref{equ:27}) that at sizes of $v$ less than the volume of a hard core of particles, of the whole distribution $p_m^v$ only two terms, $p_0^v$ and $p_1^v$ differ from zero. Since (\ref{equ:37}) gives $\overline m = \varrho v$, we have
\begin{equation}
\overline m = 0 \times p_0^v + 1 \times p_1^v = p_1^v = \varrho v. 
\label{equ:44}
\end{equation}

Then $p_0^v = 1 - \varrho v$ and 
\begin{equation}
\Upsilon_v = \frac{1}{1 - \varrho v}, 
\label{equ:45}
\end{equation}
which is a universal expression for OSE partition function at the system sizes tending to zero. 

Consideration similar to the above one may be done for the volume containing not more than $2, 3,$ etc., atoms.

It is clear that no analog may be given (except formal expressions) for GCE, since in this case the derivation of basic relations makes use of the assumption of large sizes of the system. However, if we do it, we shall obtain the analogs of (\ref{equ:44}), (\ref{equ:45}) with the replacement of $\varrho \rightarrow z$ which is obviously absurd.

Now consider fluctuations in a small volume. With this aim we return to equation (\ref{equ:43}). Let us use local character of the Ursell distribution function 
\begin{equation}
 {\cal F}^{(2)}_{1,2} = \varrho^{(2)}_{G,1,2} - \varrho^{(1)}_{G,1} \varrho^{(1)}_{G,2}. 
\label{equ:46}
\end{equation}

Transforming (\ref{equ:43}),and in view of
\begin{equation}
\int  {\cal F}^{(2)}_{1,2} d\bm{r}_2 = \varrho ( \varrho k_B T \varkappa_T - 1), 
\label{equ:47}
\end{equation}
where $\varkappa_T$ is the isothermal compressibility, we obtain
\begin{equation}
\frac{\overline {m^2} - {\overline m}^2}{\overline {m}} = \varrho k_B T \varkappa_T  - \frac{1}{\overline {m}} \int  \psi^v_1 \chi^v_2 {\cal F}^{(2)}_{1,2} d\bm{r}_1 d\bm{r}_2. 
\label{equ:48}
\end{equation}

This expression, just as the original one (\ref{equ:43}), is exact and holds for any volumes. We see that, along with ordinary term $\varrho k_B T \varkappa_T$, an additional term appears in the right-hand side.

For volumes restricted by rather smooth surfaces, and hence for rather large volumes (compared with the molecular size) the second term in the right-hand side of (\ref{equ:48}) is proportional to the surface (see Section \ref{subsec:05a}). Performing integration over the surface in (\ref{equ:48}), one has 
\begin{equation}
\frac{\overline {m^2} - {\overline m}^2}{\overline {m}} = \varrho k_B T \varkappa_T  - \frac{a}{\overline {m}} \int  \psi^v_1 \chi^v_2 {\cal F}^{(2)}_{1,2} dx_1 d\bm{r}_2, 
\label{equ:49}
\end{equation}
where $x_1$ is the coordinate perpendicular to the system surface.

Since ${\cal F}^{(2)}_{1,2}$ depends only on the relative configuration of particles, we can perform integration over $x_1$ in (\ref{equ:49}). Making the change of variables $\bm{r}_2' = \bm{r}_2 - \bm{r}_1$ and performing some other simple transformations, we get

\begin{equation}
\frac{\overline {m^2} - {\overline m}^2}{\overline {m}} = \varrho k_B T \varkappa_T  - \frac{a}{\overline {m}} \int_{x>0}  x {\cal F}^{(2)}(r) d\bm{r}. 
\label{equ:50}
\end{equation}

Equations (\ref{equ:49}), (\ref{equ:50}) give the expressions for the mean square relative deviation with allowance for surface corrections. 

For absolute form we have

\begin{equation}
\overline {m^2} - {\overline m}^2 - \overline {m} =v \int {\cal F}^{(2)}(r) d\bm{r}  - a \int_{x>0}  x {\cal F}^{(2)}(r) d\bm{r}. 
\label{equ:51}
\end{equation}

For even more small volumes where separation of surface terms is impossible other expressions are valid. Let us use the fact that it is impossible to place two atoms into the volume less than that of a particle hard core. For such a volume, it follows from (\ref{equ:42}) that for $k > 1$
\begin{equation}
\left \langle  \frac{m!}{(m-k)!} \right \rangle = 0. 
\label{equ:52}
\end{equation}
For example, for $k=2$, we obtain
\begin{equation}
\overline {m^2} - \overline {m} = 0 
\label{equ:53}
\end{equation}
and
\begin{equation}
\frac{\overline {m^2} - {\overline m}^2}{\overline {m}} = 1 - \varrho v.
\label{equ:54}
\end{equation}

Using (\ref{equ:43}), we can also get the exact expressions. Performing a procedure similar to the transition from (\ref{equ:49}) to (\ref{equ:50}), we obtain

\begin{equation}
\int  f(\bm{r})  {\cal F}^{(2)}(r) d\bm{r} =  \overline {m^2} - {\overline m}^2 - \overline {m},
\label{equ:55}
\end{equation}

where

\begin{equation}
f(\bm{r}) = \int \psi^v(\bm{r}_1) \psi^v(\bm{r}_1 + \bm{r}) d\bm{r}_1.
\label{equ:56}
\end{equation}

Recall that $\psi^v$ is a characteristic function, specifying a certain volume. Equality (\ref {equ:55}) holds for the volume of any shape and size. For example, in the case of the sphere, we have

\begin{equation}
f(\bm{r}) = 
	 \left\{ 
			\begin{array}{ll} 
         \pi(\frac{4}{3}R^3 - R^2 r + \frac{1}{12} r^3) & (r < 2R)\\   
         0 & (r \geq 2R)
     	\end{array}  
		\right.
		\label{equ:57},
\end{equation}

where $R$ is the radius of the sphere. Substituting $f$ in (\ref{equ:55}), we obtain the exact expression

\begin{eqnarray}
&& \overline {m^2} - {\overline m}^2 - \overline {m} = \frac{4}{3}\pi R^3 \int \limits_{0}^{2R}  {\cal F}^{(2)}(r) d\bm{r} \nonumber \\
&&  -\pi R^2 \int \limits_{0}^{2R} r {\cal F}^{(2)}(r) d\bm{r} + \frac{\pi}{12} \int \limits_{0}^{2R} r^3 {\cal F}^{(2)}(r) d\bm{r}.
\label{equ:58}
\end{eqnarray}

When the sphere radius $R$ exceeds the correlation radius (the radius of ${\cal F}^{(2)}(r)$ oscillations decay), the first term in the right side of (\ref{equ:58}) coincides with that in the right-hand side of (\ref{equ:51}), and the second - with the second one, respectively, as can be verified by performing the integration over the angles in the last term of equation (\ref{equ:51}).

Thus, the surface terms of the mean-square deviations are determined by the first moments of the pair Ursell function or, eventually, by the first moments of the radial distribution function.

If $R \rightarrow 0$, ${\cal F}^{(2)}(r) \rightarrow - \varrho^2$ and using (\ref{equ:58}) we obtain

\begin{equation}
\overline {m^2} - {\overline m}^2 - \overline {m} = -\frac{16}{9} \pi^2 R^6 \varrho^2,
\label{equ:59}
\end{equation}

which coincides with (\ref{equ:53}).

Higher order fluctuations and fluctuations for the volumes of greater sizes may be calculated in a similar way.

\subsection{\label{subsec:04e}Scale invariance} 

Probabilistic distributions for two different volumes are obviously connected by complete probability formula
\begin{equation}
w^v_m = \sum_{N=m}^\infty P(m|N) W^V_N,
\label{equ:60}
\end{equation}
where $w^v_m$ is the unconditional probability to find $m$ particles in the volume $v$, $W^V_N$ is the unconditional probability to find $N$ particles in the volume $V$, and $P(m|N)$ is the probability to find $m$ particles in the volume $v$ on condition that the volume $V$ contains $N$ particles. 

Naturally the identity of intensive parameters of the medium and its homogeneity are assumed. Consider the case where the corresponding volumes are inserted in one another, and their boundaries are separated by a sufficient distance.

So statistical distribution scale invariance is to mean that the following equality holds
\begin{equation}
w^V_N = W^V_N.
\label{equ:61}
\end{equation}

In other words, functional dependences of unconditional probabilities are to be the same for different volumes.

Comparing (\ref{equ:60}) with (\ref{equ:25}), we arrive at the conclusion that scale invariance is absent in GCE, since, though the character of relation (\ref{equ:25}) is similar to (\ref{equ:60}), scaling condition (\ref{equ:61}) is violated; this is seen from the comparison between (\ref{equ:06}) and (\ref{equ:27}). Note that the volume term retains its form; therefore, this part of GCE is scaled.

To prove OSE scale invariance, first it is necessary to construct the conditional probability $P(m|N)$. We cannot use $p^v_{C,m}$ (\ref{equ:24}) because of the additional condition: there is a surface at the boundary of the volume $V$.

To find $P(m|N)$, employ multiplication rule
\begin{equation}
P(m \cap N) =  P(m|N) p^V_N,
\label{equ:62}
\end{equation}
where $P(m \cap N)$ is the probability of a simultaneous finding of $m$ particles in the volume $v$ and $N$ particles in the volume $V$. Calculation of this quantity by analogy with transformations (\ref{equ:24}) $\rightarrow$ (\ref{equ:26}) yields
\begin{widetext}
%\fontsize{10pt}{12pt} \selectfont
\begin{equation}
 P(m \cap N) =  \frac{z^N}{m!(N-m)!\Xi_{V'}}\sum_{M=0}^\infty \frac{z^M}{M!}\int\exp(-\beta U^{N+M}_{1...N+M}) \left [\prod_{i = 1}^m\prod_{j = m+1}^N\prod_{k = N+1}^{N+M}\psi^v_i (\psi^V_j-\psi^v_j)(\psi^{V'}_k-\psi^V_k) \right ]  d\bm{r}_1...d\bm{r}_{N+M}, \label{equ:63}
\end{equation}
\end{widetext}
%\fontsize{12pt}{14pt} \selectfont
where $V'$ is some volume exceeding $V$. Using the binomial formula and the Boltzmann factor symmetry under particle permutation again, it can readily be seen that 
\begin{equation}
p^v_m = \sum_{N=m}^\infty P(m \cap N).
\label{equ:64}
\end{equation}

To make calculations we change the summation order again, and  $p^v_m$ is defined by the expression of (\ref{equ:27}) type with the replacement of $V \rightarrow V'$.

Besides, it is easily established that
\begin{equation}
p^V_N = \sum_{m=0}^N P(m \cap N).
\label{equ:65}
\end{equation}

For the desired conditional probability, we obtain from (\ref{equ:62})
\begin{widetext}
\begin{equation}
 P(m|N) =\binom{N}{m}\frac{\displaystyle \sum_{M=0}^\infty \frac{z^M}{M!}\int\exp(-\beta U^{N+M}_{1...N+M})  \left [\prod_{i = 1}^m\prod_{j = m+1}^N\prod_{k = N+1}^{N+M}\psi^v_i (\psi^V_j-\psi^v_j)(\psi^{V'}_k-\psi^V_k) \right ]  d\bm{r}_1...d\bm{r}_{N+M}}    {\displaystyle \sum_{M=0}^\infty \frac{z^M}{M!}\int\exp(-\beta U^{N+M}_{1...N+M})  \left [\prod_{i = 1}^N\prod_{k = N+1}^{N+M}\psi^V_i (\psi^{V'}_k-\psi^V_k) \right ]  d\bm{r}_1...d\bm{r}_{N+M}}.
 \label{equ:66}
\end{equation}
\end{widetext}

As is seen, after division the first term of series in powers of $z$ (\ref{equ:66}) coincides with expression (\ref{equ:24}).

Eventually, based on properties (\ref{equ:64}), (\ref{equ:65}) and the form of expression (\ref{equ:66}) one can write
\begin{equation}
p^v_m = \sum_{N=m}^\infty P(m|N) p^V_N
\label{equ:67}
\end{equation}
and thus OSE scale invariance is proved.

\section{\label{sec:05}Series in activity}

In this section we give the expressions for GCE and OSE objects as the series in powers of activity $z$, which are analogues to the Mayer series for dense gases. The first subsection presents the separation procedure for volume and surface terms. Finally, the convergence of series will be discussed. 

Here we assume that the systems under discussion are restricted by rather smooth surfaces, i.e., they are deprived of asperities of molecular sizes. Consequently, this condition imposes the constraints on the systems size as well, in the cases where it is necessary - for OSE distributions it must essentially exceed molecular sizes.

\subsection{\label{subsec:05a}The series for GCE and the procedure of separating surface terms}

Taking the logarithm of equality  (\ref{equ:08}), in view of (\ref{equ:04}) and the logarithmic form of generating function for Ursell factors ${\cal U}^{(t)}_{1...t}$ (see Appendix \ref{subsec:appenda1}), one has 
\begin{equation}
\ln{\Xi_V} =  \sum_{t=1}^\infty \frac{z^t}{t!} \int \left [ \prod_{i = 1}^{t} \psi^V_i \right ]  {\cal U}^{(t)}_{1...t} d\bm{r}_1...d\bm{r}_t.
\label{equ:68}  
\end{equation}

Transforming (\ref{equ:68}) by adding and subtraction of identical terms, we obtain

\begin{eqnarray}
\ln{\Xi_V} &=& \sum_{t=1}^\infty \frac{z^t}{t!} \int \psi^V_1 {\cal U}^{(t)}_{1...t} d\bm{r}_1...d\bm{r}_t \label{equ:69} \\
&-& \sum_{t=2}^\infty \frac{z^t}{t!} \int \psi^V_1 \left [1 - \prod_{i = 2}^{t} (1 - \chi^V_i) \right ] {\cal U}^{(t)}_{1...t} d\bm{r}_1...d\bm{r}_t.\nonumber
\end{eqnarray}

The first term in the right-hand side of (\ref{equ:69}), the volume term, is the Mayer familiar expansion
\begin{equation}
Pv = k_BT\sum_{t=1}^\infty \frac{z^t}{t!} \int \psi^v_1{\cal U}^{(t)}_{1...t} d\bm{r}_1...d\bm{r}_t 
\label{equ:70}  
\end{equation}
or
\begin{equation}
P(z,T) =zk_BT  + k_BT \sum_{k=2}^\infty \frac{z^k}{k!} \int {\cal U}^{(k)}_{1...k} d\bm{r}_2...d\bm{r}_k.
\label{equ:71} 
\end{equation}

However, it has the analog for surface tension in the form given below.

The second term in the right-hand side of (\ref{equ:69}) is proportional to the surface size. Indeed, its structure is such that at least two particles always reside on different sides of the boundary. If we expand the product in it, all terms will have the form
\begin{equation}
\int \psi^V_1 \left [ \prod_{i = 2}^{j} \chi^V_i \right ] {\cal U}^{(t)}_{1...t} d\bm{r}_1...d\bm{r}_t,
\label{equ:72}  
\end{equation}
where $2 \leq j \leq t $. For the fixed first particle, by virtue of Ursell factors locality, the integrals of the type of (\ref{equ:72}) are defined by local region near it. So $\int ... d\bm{r}_2...d\bm{r}_t$  do not depend on the first particle displacement along the system boundary. When the first particle moves away from the boundary, they decay rapidly due to the fixing factors  $\chi^V_i$ and Ursell factors locality. Performing integration over the surface and placing the area outside the summation sign, we arrive at (\ref{equ:23}), where $A\sigma (z,T)$ has the form
\begin{eqnarray}
A\sigma (z,T) &=& k_BT\sum_{t=2}^\infty \frac{z^t}{t!} \label{equ:73} \\
& \times &  \int \psi^V_1 \left [1 - \prod_{i = 2}^{t} (1 - \chi^V_i) \right ] {\cal U}^{(t)}_{1...t} d\bm{r}_1 ...d\bm{r}_t,\nonumber
\end{eqnarray}
and for $\sigma (z,T)$ we have the expression
\begin{eqnarray}
\sigma (z,T) &=& k_BT\sum_{t=2}^\infty \frac{z^t}{t!} \label{equ:74} \\
&\times &  \int \psi^V_1 \left [1 - \prod_{i = 2}^{t} \psi^V_i \right ] {\cal U}^{(t)}_{1...t} dx_1 d\bm{r}_2 ...d\bm{r}_t,\nonumber
\end{eqnarray}
where $x_1$ is the coordinate perpendicular to the surface defined by the boundary $\psi^V_1$. The axis $x_1$ direction is chosen such that $dx_1 > 0$.

The first terms of expression (\ref{equ:74}) were obtained in paper \cite{Bellemans1962} by diagram technique; however, construction of the far terms of the series in the same way seems extremely difficult.

The translational invariance of ${\cal U}^{(t)}_{1 ... t}$ permits to perform the integration over the $x_1$, just as we did when considering the fluctuations in small systems (section \ref{subsec:04d}). 

Performing  the change of variables $\bm{r} _i'= \bm{r}_i - \bm{r}_1$ in (\ref{equ:74}), shifting the coordinates of the Ursell factors, and integrating over $ x_1 $, arrive at

\begin{eqnarray}
\sigma (z,T) =&& k_BT\sum_{t=2}^\infty \frac{z^t}{t(t-2)!} \nonumber \\
&& \times \int_{ x_2> \{0, x_3... x_k \} }  x_2  {\cal U}^{(t)}_{0,2...t}  d\bm{r}_2 ...d\bm{r}_t,
\label{equ:75}
\end{eqnarray}

where we use the Ursell factors symmetry under particle permutations.

Finally, using the change of variables $\bm{r}'_i = \bm{r}_i - \bm{r}_2$, for $i = 3,...t$, we have
\begin{eqnarray}
\nu (z,T) =&& k_BT\sum_{t=2}^\infty \frac{z^t}{t(t-2)!} \nonumber \\
&& \times \int_{ x_2 > 0,\dots x_t > 0}  x_2  {\cal U}^{(t)}_{0,2...t}  d\bm{r}_2 ...d\bm{r}_t.
\label{equ:76}
\end{eqnarray}

In this case we used the invariance of ${\cal U}^{(t)}$ under particle permutations and under spatial inversion ($\bm{r} _i'= - \bm{r}_i$).

\subsection{\label{subsec:05b}Series for OSE}

Taking the logarithm of both sides of equation (\ref{equ:18}) and passing to the limit $V \to \infty$, we obtain 

\begin{equation}
\ln{\Upsilon_v} =  \lim_{V\to \infty} \left ( \ln{\Xi_V} - \ln{\Xi_{V-v}} \right ),
\label{equ:77}  
\end{equation}
of course, if the intensive parameters of the environment and the parameters of the volume $v$ are preserved.

Substitution of (\ref{equ:68}) gives

\begin{eqnarray}
\ln{\Upsilon_v} &=& \lim_{V\to \infty} \sum_{t=1}^\infty \frac{z^t}{t!}\int {\cal U}^{(t)}_{1...t} \label{equ:78} \\
&&\times   \left [ \prod_{i = 1}^{t} \psi^V_i  - \prod_{i = 1}^{t} (\psi^V_i - \psi^v_i) \right ]  d\bm{r}_1...d\bm{r}_t.\nonumber
\end{eqnarray}

Using local character of Ursell factors (\ref{equ:a02}), it can easily be shown that for OSE partition function we have the series
\begin{equation}
\ln{\Upsilon_v} = \sum_{t=1}^\infty \frac{z^t}{t!} \int \left [1 - \prod_{i = 1}^{t} \chi^v_i \right ] {\cal U}^{(t)}_{1...t} d\bm{r}_1...d\bm{r}_t.
\label{equ:79} 
\end{equation}

It is interesting to compare it with the expression for GCE partition function logarithm (\ref{equ:68}). 

Separating volume and surface terms by the above technique, we obtain (\ref{equ:21}) and (\ref{equ:74}) with the replacement of $\psi^v_i \leftrightarrow \chi^v_i$ which is always possible due to the spatial symmetry of the problem.

To derive the series for the general term of OSE distribution we apply the expression for distribution function (\ref{equ:a05}). Substituting it in (\ref{equ:27}) gives

\begin{eqnarray}
p^v_m &=& \frac{z^m}{m! \Upsilon_v}\sum_{t=0}^\infty \frac{z^t}{t!} \label{equ:80} \\
&\times& \int \left [ \prod_{i = 1}^m \prod_{j = m+1}^{m+t}\psi^v_i\chi^v_j \right ] {\cal B}^{(m,t)}_{1...m+t} d\bm{r}_1...d\bm{r}_{m+t}, \nonumber
\end{eqnarray}
where ${\cal B}^{(m,k)}_{1...m+k}$ are partial localization factors intermediate between Boltzmann and Ursell factors (see Appendix \ref{subsec:appenda2}).

As is easily seen the first term of series (\ref{equ:80}) coincides with GCE distribution (\ref{equ:06}) by virtue of the equality ${\cal B}^{(m,0)}_{1...m} = \exp(-\beta U^{m}_{1...m})$ (\ref{equ:a06}) accurate to normalizing factors - partition functions.
 
The structure of series (\ref{equ:80}) is of interest. The position of delocalized group of $m$ particles is defined by characteristic functions $\psi^v_i$ inside the system volume. The position of the localized group of $t$ particles outside it - by the functions $\chi^v_j$, but in close relation to the volume owing to functions ${\cal B}^{(m,t)}_{1...m+t}$. Summation is made over greater and greater clusters. One can say that volume properties of the distribution are specified by the first $m$ particles - ordinary, delocalized, while surface ones - by  $t$ particles - localized.

Since the distinctions between GCE and OSE are related to surface terms, therefore, all terms of series (\ref{equ:80}) beginning with the 2-d term contain both the volume and the surface of the system in different powers.

\subsection{\label{subsec:05c}Convergence of series}

Convergence of series in powers of activity where the integrals of Ursell factors and partial localization factors are used as the expansion coefficients is studied, in particular, in paper \cite{ruellestatmeh1969}. References to other original works are also given in this paper. 

Summarizing the results, one can say that these series converge at least for dense gases where the convergence radius is defined by the interaction potential. In a great number of cases the convergence condition is of the form 
\begin{equation}
z \int {\cal U}^{(2)}_{1,2} d\bm{r}_2 \leq 1,
\label{equ:81} 
\end{equation}
or something like that. The potentials with "hard" core - hard spheres or Lennard-Jones potentials (the so-called stable and regular potentials) satisfy the conditions providing the convergence of such series. Of course, the Coulomb potentials lead to divergence on the infinity. 

We can conclude that at least for such potentials and conditions of (\ref{equ:81}) type the series considered in this section converge.

\section{\label{sec:06}Discussion}

\subsection{\label{subsec:06a}OSE-limit}

In (\ref{equ:79}), (\ref{equ:80}) passing to OSE-limit is performed
\begin{equation} 
  \begin{minipage}[c]{0.80\linewidth}
    \centering
    $V \rightarrow \infty$~~~  \\
    $v = const$  \\
    $ip = const$,
    \end{minipage}
  \label{equ:82} 
\end{equation}
where $ip$ are intensive parameters of the medium. Expression (\ref{equ:27}) takes the form
\begin{equation}
p^v_m = \frac{1}{m! \Upsilon_v} \int \left [ \prod_{i = 1}^m \psi^v_i \right ] \varrho^{(m)}_{1...m}(\chi^v) d\bm{r}_1...d\bm{r}_m. 
\label{equ:83}
\end{equation}

Here $\varrho^{(m)}_{1...m}$ has the meaning of the probability density of finding some configuration of $m$ particles in the given volume $v$, on condition that all other particles are outside it. This probability density is defined for infinite homogeneous medium, and has the series of (\ref{equ:a05}) type as one of representations in the cases where the series converges. 

The thermodynamic limit in the form of (\ref{equ:82}) is preferable to the limit commonly used for GCE
\begin{equation} 
  \begin{minipage}[c]{0.80\linewidth}
    \centering
    $V \rightarrow \infty$~~~  \\
    $ip = const$.
    \end{minipage}
  \label{equ:84} 
\end{equation}

The thing is that in the last case, despite the increased size of the system, the walls of the vessel are always "stuck" to the system boundaries; this leads to surface deformation preservation and violation of the relation of the mean number of particles. This will be the subject of discussion in the next section.

\subsection{\label{subsec:06b}The difference between OSE and GCE}

For illustration in this section we consider two ensembles - GCE and OSE that are of the same volume $V$.

Differentiating (\ref{equ:23}), (\ref{equ:21}) with respect to chemical potential, we have
\begin{equation}
\frac{\partial\ln{\Xi_V}}{\partial \ln{z}} = N_b + N_s = \overline{N}
\label{equ:85}
\end{equation} 
and
\begin{equation}
\frac{\partial\ln{\Upsilon_V}}{\partial \ln{z}} = N_b - N_s = \overline{N} - N_s,
\label{equ:86}
\end{equation} 
where
\begin{equation}
N_b = V\frac{\partial P}{\partial \mu}
\label{equ:87}
\end{equation}  
- the number of volume particles, and
\begin{equation}
N_s = - A\frac{\partial \sigma}{\partial \mu}
\label{equ:88}
\end{equation}
- that of surface ones.

As is known, the last equality in (\ref{equ:85}) is readily obtained by differentiation (\ref{equ:08}) with respect to chemical potential, and in (\ref{equ:86}) we used expression (\ref{equ:38}).

Thus OSE and GCE distributions contain surface terms of different signs. The opposite sign in the surface term in (\ref{equ:23}) is not accidental; GCE, as a set of closed systems, fails to completely reproduce an open system.

Since each of the terms of series (\ref{equ:08}) involves the surface corresponding to a closed system, GCE is a hybrid. This ensemble is open from the standpoint of the volume properties, because it describes adequately the fluctuations of the total number of particles, and is closed from the point of view of the surface properties that is obvious from (\ref{equ:85}). Probably, it is not a coincidence that the author of paper \cite{Bellemans1962} has concluded that GCE describes a "drop". It should only be added that it is a "drop" restricted by potential barriers.

Now we can return to equation (\ref{equ:11}). It follows from (\ref{equ:11}) and (\ref{equ:85}) that
\begin{equation}
\int \psi^V_1 \varrho^{(1)}_{G,1} (\psi^V) d\bm{r}_1 = \overline{N} = N_b + N_s.
\label{equ:89}
\end{equation}

So we see that $\varrho^{(k)}_{G,1...k}$ involve surface terms, and (\ref{equ:89}) does not fit the conception of the GCE as part of a homogeneous medium.

For comparison, we derive from (\ref{equ:42}) and (\ref{equ:86}) for OSE
\begin{equation}
\int \psi^V_1 \varrho^{(1)}_{1} d\bm{r}_1 = \overline{N} = N_b, 
\label{equ:90}
\end{equation}
which is related to the absence of near surface density distortions in the given case.

\subsection{\label{subsec:06c}Small fluctuations}

It should be emphasized once again that in the case under study the surface terms appear in the description of a homogeneous medium involving no real surface but just a hypothetical one restricting the separated system. This unexpected result becomes clearer taking into account that in statistical distributions only fluctuations within the limits of the separated volume are of interest to us. The states corresponding to these fluctuations already contain the restricting surface, and this gives rise to the above effect. 

Consequently, the surface terms are maximum for great fluctuations, and are compensated for small values. Really, it would be strange if the probability of finding the mean number of particles in a certain volume contained the surface. Such compensation actually takes place.

First this is evident from the equality $m_b = \overline m$ itself. Since at small values of the number of particles in a given volume the probability depends on the value of the restricting surface, as is clear from (\ref{equ:29}), and the mean value is independent of it, the only variant is compensation of the surface quantities for distribution terms with large $m$.

Second, calculations show that for the first expansion terms in activity we have a rigorous compensation of surface terms near the mean values. 

Finally, for the general case it can easily be shown by considering OSE distribution near the mean values that the above compensation holds, and distribution terms corresponding to the mean values do not have surface components.

\subsection{\label{subsec:06d}Surface particles}

Since $\sigma$, according to (\ref{equ:74}), depends on two variables, we obtain the non-zero value of the number of surface particles
\begin{equation}
N_S = - A \frac{\partial \sigma}{\partial \mu} = - \beta A z \frac{\partial \sigma}{\partial z},
\label{equ:91}
\end{equation}
or, using (\ref{equ:73}),
\begin{equation}
N_S = \sum_{t=2}^\infty \frac{z^t}{(t-1)!} \int \psi^V_1 \left [\prod_{i = 2}^{t} \psi^V_i - 1 \right ] {\cal U}^{(t)}_{1...t} d\bm{r}_1 ...d\bm{r}_t.
\label{equ:92}
\end{equation}

Equality (\ref{equ:92}) is almost evident taking into account the well-known formula for density expansion into a series in powers of activity for GCE
\begin{equation}
\varrho_G(\bm{r}_{1},z) = z + z\sum_{n=1}^\infty \frac{z^n}{n!} \int \left [ \prod_{i = 2}^{n+1} \psi^V_i \right ]{\cal U}^{(n+1)}_{1...n+1} d\bm{r}_{2}...d\bm{r}_{n+1},
\label{equ:93}
\end{equation}
following, for example, from (\ref{equ:85}) or (\ref{equ:a05}) and analogous quantity for OSE
\begin{equation}
\varrho(z) = z + z\sum_{n=1}^\infty \frac{z^n}{n!} \int {\cal U}^{(n+1)}_{1...n+1} d\bm{r}_{2}...d\bm{r}_{n+1},
\label{equ:94}
\end{equation}
corresponding to (\ref{equ:40}), (\ref{equ:71}). The number of surface particles is identified with the quantity
\begin{equation}
N_S = \int \limits_V { \left [\varrho_G(\bm{r},z) - \varrho(z) \right ] d \bm{r}},
\label{equ:95}
\end{equation}
and the surface number density $\varrho_S = N_S/A$ is defined by the expression
\begin{equation}
\varrho_S = \int \limits_{L_t} { \left [\varrho_G(x,z) - \varrho(z) \right ] d x},
\label{equ:96}
\end{equation}
where $x$ is the coordinate perpendicular to the surface, as before; $L_t$ is the transition region near the surface.
 
In  (\ref{equ:95}), (\ref{equ:96}) we use the expression for the near surface density of particles at the outer GCE boundary. However, it is easily seen that for hard solid immersed in the system the conclusion is the same. 

Formula (\ref{equ:95}) coincides with the well-known expression for the surface number density at a impenetrable wall \cite{SokolowskiStecki1980, HendersonSwol1984}.

So, ``surface tension'' corresponds exactly to near surface (at hard solid or cavity) oscillations (deviations) of the number density.

\section{\label{sec:07}Summary}

\begin{enumerate}
	\item A new ensemble - OSE (open statistical ensemble) is presented, the main peculiarity of which is a correct account of surface terms for an open system. This results in the replacement of the Bolzmann factors in configuration integrals of a grand canonical ensemble (GCE) by distribution functions of a specific type (\ref{equ:27}).	
	\item OSE partion function involves volume and surface terms (\ref{equ:21}), and in the expanded form is given by expression (\ref{equ:79}). Similar expression for OSE distribution is given by sum (\ref{equ:80}), with GCE being the first term up to a factor.
	\item The ``surface tension coefficient'' involved in OSE partition function corresponds to non-zero number of surface particles which is unambiguously determined by the deviations of the number density from the mean value near fluctuation cavities (\ref{equ:95}).	
	\item The expression for OSE partition function agrees with the thermodynamic approach for fluctuation formation probability (\ref{equ:30}).	
	\item In contrast to GCE, OSE has the property of scale invariance: distributions in the initial and embedded volumes functionally coincide (\ref{equ:67}).	
	\item OSE distribution has no lower bound of the volume, and may be directly applied to the studies of small systems (Section \ref{subsec:04d}).	
	\item Thermodynamic and statistical OSE relations for dense gases are provided by recurrence relations (Appendix \ref{sec:appendb}) of a new class of functions ${\cal B}^{(m,k)}_{1...m+k}$ (Appendix \ref{subsec:appenda2}).		
	\item The functions ${\cal B}^{(m,k)}_{1...m+k}$ (\ref{equ:a04}) generalize the concepts of Boltzmann and Ursell factors (\ref{equ:a01}), and involve them as the extreme values (\ref{equ:a06}), (\ref{equ:a07}).	
	\item The basic expressions and properties of OSE may be obtained both on the level of factors, and on the level of distribution functions. Thus problems related to the convergence of series in number density (activity) may be avoided. For example, there are two types of expressions for the general term of OSE distribution (\ref{equ:80}), (\ref{equ:83}).	
	\item For a homogeneous medium the mean number of particles for OSE, unlike GCE, involves no surface terms: $\overline m = m_b$. 
	\item GCE is a hybrid the volume terms of which correspond to an open system, and the surface ones - to a closed system (Section \ref{subsec:06b}). In particular, GCE partition function contains surface terms with the sign that is opposite to OSE one (\ref{equ:21}), and corresponds to a closed system (\ref{equ:23}).	
	
\end{enumerate}

\appendix

\section{\label{sec:appenda}Factors}

\subsection{\label{subsec:appenda1}Ursell factors}

Ursell factors  ${\cal U}^{(k)}_{1...k}$ are also called cluster functions. Just these factors appear in the Mayer well-known expansion in powers of activity for pressure (\ref{equ:71}) \cite[p.129]{hillstatmeh1987}, \cite[p.232]{landaulifshitz1985}. (From this it follows that a logarithm serves as a generating function for Ursell factors). First they were introduced in paper \cite{ursell1927}. 

They may be defined by the equality 
\begin{eqnarray}
{\cal U}^{(k)}_{1...k} &=& \sum_{\{\bm{n}\}}(-1)^{l-1}(l-1)!\prod_{\alpha = 1}^l \exp(-\beta U^{k_\alpha}(\{\bm{n}_\alpha\})),  \nonumber \\
1 & \leq & ~ k_\alpha \leq k, ~~ \sum_{\alpha = 1}^l k_\alpha = k, ~~ \exp(-\beta U^{1}) = 1,
\label{equ:a01}
\end{eqnarray}
where $\{\bm{n}\}$ denotes some partition of the given set of $k$ particles with the coordinates $\bm{r}_1,...\bm{r}_k$ into disjoint groups $\{\bm{n}_\alpha\}$, $l$ is the number of groups of a particular partition, $k_\alpha$ is the size of the group with the number $\alpha$, the sum is taken over all possible partitions, and the meaning of the condition $\exp(-\beta U^{1}) = 1$ is apparent: single groups make no contribution into the products.

For illustration, several first ${\cal U}^{(k)}_{1...k}$ are given
\begin{eqnarray}
{\cal U}^{(1)}_{1}\mkern 9mu &=& 1 \label{equ:a02} \\
{\cal U}^{(2)}_{1,2} \mkern 8mu &=& \exp(-\beta U^{2}_{1,2}) - 1 \nonumber \\
{\cal U}^{(3)}_{1,2,3} &=& \exp(-\beta U^{3}_{1,2,3}) - \exp(-\beta U^{2}_{1,2}) \nonumber \\
 &-& \exp(-\beta U^{2}_{1,3}) - \exp(-\beta U^{2}_{2,3}) + 2 \nonumber \\
\dotso \nonumber
\end{eqnarray}
Ursell factors decay rapidly when any group of particles (including single one) moves away.

\subsection{\label{subsec:appenda2}Partial localization factors}

These quantities are a hybrid of Boltzmann and Ursell factors. They play an important role in OSE mathematical formalism. As far as we know, they were introduced for the first time in paper \cite{ruellestatmeh1969}. 

A portion of particles appearing in these functions do not cause the decay when they move away (delocalized group), another portion does it (localized one).

Introduce the designation for them
\begin{equation}
{\cal B}^{(m,k)}_{1...m+k}.
\label{equ:a03}
\end{equation}

Here the superscripts $m$ and $k$ define the number of delocalized and localized particles, respectively ($m = 1,2,3,\dots, k = 0,1,2,\dots$). The subscripts denote the coordinates of particles. Let us take that the first $m$ particles are delocalized, and other particles are localized. 

The structure of these functions is similar to that of Ursell factors of the $k+1$ rank, however, note that in construction by (\ref{equ:a01}) type, the first $m$ particles (delocalized) are treated as one compound particle. In other words, define ${\cal B}^{(m,k)}_{1...m+k}$ by the equality 
\begin{eqnarray}
{\cal B}^{(m,k)}_{1...m+k} &=& \sum_{\{\bm{n}\}}(-1)^{l-1}(l-1)!\label{equ:a04}\\
&\times& \prod_{\alpha = 1}^l \exp(-\beta U^{k_\alpha+(m-1)\delta_{\alpha\nu}}(\{\bm{n}_\alpha\})),  \nonumber \\
1  \leq&  k_\alpha& \leq k+1; ~~ \sum_{\alpha = 1}^l k_\alpha = k+1; ~~ \exp(-\beta U^{1}) = 1,\nonumber 
\end{eqnarray}
where the designations are analogous to (\ref{equ:a01}) on condition that the sum is taken over all possible partitions of the set of $k+1$ particles including one compound particle. $\delta_{\alpha\nu}$ is the Kroneker delta, and $\nu$ is the group number, which involves a compound particle.

The generating function for ${\cal B}^{(m,k)}_{1...m+k}$  is distribution function of GCE type  $\varrho^{(m)}_{G,1...m}$. Expanding the partition function $\Xi_V$ in (\ref{equ:10}), and dividing the series, we have
\begin{eqnarray}
&&\varrho^{(m)}_{G,1...m} (\psi^V) = z^m \left \{ {\cal B}^{(m,0)}_{1...m}  +  \sum_{k=1}^\infty \frac{z^k}{k!} \right . \label{equ:a05} \\
&& ~~~~~~ \left . \times  \int   \left [ \prod_{i = m+1}^{m+k} \psi^V_i \right ]   {\cal B}^{(m,k)}_{1...m+k} d\bm{r}_{m+1}...d\bm{r}_{m+k} \right \}.  \nonumber
\end{eqnarray}

The proof of this relation is given in Appendix \ref{subsec:appendb1}.

${\cal B}^{(m,k)}_{1...m+k}$ first in the localized group are 
\begin{eqnarray}
{\cal B}^{(m,0)}_{1...m} \mkern 12mu &=& \exp(-\beta U^{m}_{1...m})\label{equ:a06} \\
{\cal B}^{(m,1)}_{1...m+1} &=& \exp(-\beta U^{m+1}_{1...m+1}) - \exp(-\beta U^{m}_{1...m})\nonumber \\
{\cal B}^{(m,2)}_{1...m+2} & = & \exp(-\beta U^{m+2}_{1...m+2}) -  \exp(-\beta U^{m+1}_{1...m+1})\nonumber \\
 &-& \exp(-\beta U^{m+1}_{1...m,m+2}) - \exp(-\beta U^{m}_{1...m}) \nonumber \\
 &\times& \exp(-\beta U^{2}_{m+1,m+2})  + 2\exp(-\beta U^{m}_{1...m}) \nonumber \\
\dotso \nonumber
\end{eqnarray} 
 
and in the delocalized one -
\begin{equation}
{\cal B}^{(1,k-1)}_{1...k} = {\cal U}^{(k)}_{1...k}
\label{equ:a07},
\end{equation}
including the case where $k=1$ (for homogeneous medium) 
\begin{equation}
{\cal B}^{(1,0)}_{1} = 1
\label{equ:a08}.
\end{equation}

As is seen from (\ref{equ:a06}), (\ref{equ:a07}), partial localization factors generalize the notions of Boltzmann and Ursell factors including them as the limiting cases.

For the factors ${\cal B}^{(m,k)}_{1...m+k}$ a number of recurrence relations are valid (see Appendix \ref{sec:appendb}) that provide the fulfillment of various physical relations.

\section{\label{sec:appendb}Recurrence relations for ${\cal B}^{(m,k)}_{1...m+k}$}

So, each operation with OSE distribution is ensured by a definite class of recurrence relations for ${\cal B}^{(m,k)}_{1...m+k}$. For brevity, we shall say that operation generates a recurrence relation or a class. Since operations considered below have already been proved by alternative methods, this means that the recurrence relations generated by them are also proved. Let us examine some of these operations.

\subsection{\label{subsec:appendb1}Correspondence to the definition}

Substituting expression (\ref{equ:10}) in (\ref{equ:a05}), expanding the series for $\Xi_V$ and performing multiplication of the series, we obtain
\begin{equation}
{\cal B}^{(m,k)}_{1...m+k} = {\cal B}^{(m+k,0)}_{1...m+k}  - \sum_{n=1}^k \sum_{perm} {\cal B}^{(n,0)}_{1...n}{\cal B}^{(m,k-n)}_{n+1...m+k}, \label{equ:b01}
\end{equation}
where $m \geq 1$, $k \geq 1$, and the internal sum is taken over samplings of localized particles only ($n$ from $k$). Relation (\ref{equ:b01}) is proved either by direct enumeration of partitions according to (\ref{equ:a04}), or by repeat substitution of the expression for ${\cal B}^{(m,k)}_{1...m+k}$ in the right-hand side of (\ref{equ:b01}).

Paper \cite{percus1964} gives another recurrence relation for Ursell functions which in terms of ${\cal B}^{(m,k)}_{1...m+k}$ looks like
\begin{equation}
{\cal B}^{(1,k)}_{1...k+1} = {\cal B}^{(k+1,0)}_{1...k+1}  - \sum_{n=1}^{k} \binom{k}{n} \left [ {\cal B}^{(n,0)}_{1...n} {\cal B}^{(1,k-n)}_{n+1...k+1}\right ]_{perm}, \label{equ:b02}
\end{equation}
where $k \geq 1$, and square brackets mean averaging over particles permutations.

In principle, this is the same equation (\ref{equ:b01}) at $m=1$ but written in a symmetric form. The question of symmetrization of all recurrence relations for ${\cal B}^{(m,k)}_{1...m+k}$ that will arise further is beyond the scope of the present contribution. However, note that in our case symmetrization is not needed: equation (\ref{equ:b01}) holds rigorously in asymmetric form as well. Nevertheless, (\ref{equ:b01}) may also be represented in a symmetric form; however, this requires that both the sum and the left-hand side of the equation be averaged over permutations.

\def\dosimm{\stackrel{sim}{=}}\def\convf{\hbox{\space \raise-2mm\hbox{$\textstyle  \bigotimes \atop \scriptstyle \omega$} \space}}

It is not improbable that all relations given below can also have asymmetric form but we shall employ a symmetric form of them. In some cases the symbol $\dosimm$ will be used. We imply that symmetrization is performed in the right- and left-hand sides of the equality over the indices where it is necessary. In the given case symmetrization will also mean, along with averaging over permutations, the aligning of indices to natural series, if necessary. This operation is due to property  (\ref{equ:a08}).

\subsection{\label{subsec:appendb2}Normalization condition}

$\sum p^v_m$ over all $m$ must be equal to 1, and we arrive at the equation
\begin{eqnarray}
\Upsilon_v &=&1 + \sum_{m=1}^\infty \frac{z^m}{m!}\sum_{t=0}^\infty \frac{z^t}{t!} \label{equ:b03} \\
&\times& \int \left [ \prod_{i = 1}^m \prod_{j = m+1}^{m+t}\psi^v_i\chi^v_j \right ] {\cal B}^{(m,t)}_{1...m+t} d\bm{r}_1...d\bm{r}_{m+t}. \nonumber
\end{eqnarray}

This is another form of OSE partition function. Here it is not easy to see expression (\ref{equ:79})! To show the identity of these formulae, we consider their logarithmic derivatives. Thus
\begin{equation}
\frac{1}{\Upsilon_v}\frac{\partial\Upsilon_v}{\partial z} = \beta \frac{\partial(Pv+\sigma a)}{\partial z},
\label{equ:b04}
\end{equation}  
and $\Upsilon_v$ in (\ref{equ:b04}) is to be defined by expression (\ref{equ:b03}), while $P v$ and $\sigma a$ - by (\ref{equ:70}) and (\ref{equ:73}), respectively, where ${\cal U}^{(t)}_{1...t}$ are replaced by ${\cal B}^{(1,t-1)}_{1...t}$, according to (\ref{equ:a07}).

To simplify calculations, we change the order of summation in the expressions. For (\ref{equ:b03}):
\begin{eqnarray}
\Upsilon_v &=&1 + \sum_{s=1}^\infty \frac{z^s}{s!}\int\sum_{m=1}^s \binom{s}{m} \label{equ:b05} \\
&& \times \left [ \prod_{i = 1}^m \prod_{j = m+1}^{s}\psi^v_i\chi^v_j \right ] {\cal B}^{(m,s-m)}_{1...s} d\bm{r}_1...d\bm{r}_{s}. \nonumber
\end{eqnarray}

Performing differentiation and multiplication of the series, and equating the expressions at different powers of $z$ and products $\psi^v_i$, we obtain the recurrence relations
\begin{eqnarray}
{\cal B}^{(m,k)}_{1...m+k} &\dosimm& {\cal B}^{(m-1,k+1)}_{1...m+k} \label{equ:b06} \\
&+& \sum_{n=0}^k \binom{k}{n}  {\cal B}^{(m-1,k-n)}_{1...m+k-n-1} {\cal B}^{(1,n)}_{m+k-n...m+k}, \nonumber
\end{eqnarray}
valid at $k \geq 0$ and $m \geq 2$, and
\begin{equation}
{\cal B}^{(1,k)}_{1...k+1} =   \sum_{m=1}^{k}  (-1)^m \binom{k}{m} \left [ {\cal B}^{(m,k-m)}_{1...k} - {\cal B}^{(m+1,k-m)}_{1...k+1}\right ]_{perm}, \label{equ:b07}
\end{equation}
where $k \geq 1$, and the class of relations having two free indices and containing low indices of localized particles
\begin{eqnarray}
\sum_{m=0}^{k}&(-1)^{m}& \binom{k}{m} {\cal B}^{(m+t+1,k-m)}_{1...t+k+1}~ \dosimm ~\sum_{l=0}^{k} \binom{k}{l}  \label{equ:b08}\\
\times &(-1)^{l}& \sum_{n=1}^{t} \binom{t}{n}  {\cal B}^{(l+n,k-l)}_{1...k+n} {\cal B}^{(1,t-n)}_{k+n+1...k+t+1} \nonumber \\
&+& \sum_{s=1}^{k} (-1)^{s} \binom{k}{s} {\cal B}^{(s,k-s)}_{1...k} {\cal B}^{(1,t)}_{k+1...k+t+1}, \nonumber
\end{eqnarray}
where $k \geq 1$, $t\geq 1$ and which at $k = 1$ reduces to (\ref{equ:b02}).

\subsection{\label{subsec:appendb3}Calculation of $\overline{m}$}

We proceed from the natural assumption $\overline{m} = m_b$ (\ref{equ:38}).

The outline of the proof is as follows. According to (\ref{equ:80}), $\overline{m}$ has the form
\begin{eqnarray}
\overline{m} &=& \frac{1}{\Upsilon_v}\sum_{m=1}^\infty\frac{z^m}{(m-1)!} \sum_{t=0}^\infty \frac{z^t}{t!} \label{equ:b09} \\
&\times& \int \left [ \prod_{i = 1}^m \prod_{j = m+1}^{m+t}\psi^v_i\chi^v_j \right ] {\cal B}^{(m,t)}_{1...m+t} d\bm{r}_1...d\bm{r}_{m+t}. \nonumber
\end{eqnarray}

Let us change the order of summation
\begin{eqnarray}
\overline{m} &=& \frac{1}{\Upsilon_v}\sum_{s=1}^\infty\frac{z^s}{(s-1)!}\int \sum_{m=1}^s \binom{s-1}{m-1} \label{equ:b10} \\
&\times&  \left [ \prod_{i = 1}^m \prod_{j = m+1}^{s}\psi^v_i\chi^v_j \right ] {\cal B}^{(m,s-m)}_{1...s} d\bm{r}_1...d\bm{r}_{s}. \nonumber
\end{eqnarray}
  
This expression must be set equal to (\ref{equ:87}), where  (\ref{equ:70}) should be used for $P$. Multiplying the series and equating the expressions at different powers of $z$ and the products $\psi^v_i$, we get a class of recurrence relations containing low indices of delocalized particles 
\begin{eqnarray}
\sum_{m=0}^{k}(-1)^{m}&& \binom{k}{m} {\cal B}^{(m+1,t+k-m)}_{1...t+k+1} ~ \dosimm ~ \sum_{l=1}^{k} (-1)^{l}  \label{equ:b11}\\
\times && \binom{k}{l}\sum_{n=0}^{t} \binom{t}{n}  {\cal B}^{(l,k+n-l)}_{1...k+n} {\cal B}^{(1,t-n)}_{k+n+1...k+t+1}, \nonumber
\end{eqnarray}
where $k \geq 1$, $t\geq 0$,  which at $k = 1$ reduces to (\ref{equ:b06}) with $m=2$, and at $t = 0$ - to (\ref{equ:b07}).

\tiny 
\raggedleft
VZ, 12.08.2011, v82.

\end{document}